\documentclass[final,1p,times]{elsarticle} 

\usepackage{graphicx}
\usepackage{amssymb} 
\usepackage{amsthm} 
\usepackage{lineno}

\journal{Nuclear Physics A} 

\begin{document}

\begin{frontmatter} 

\title{Anisotropic flow measured from multi-particle azimuthal correlations for Pb-Pb collisions at 2.76 TeV by ALICE at the LHC}

\author{Ante Bilandzic (for the ALICE\fnref{col1} Collaboration)}
\fntext[col1] {A list of members of the ALICE Collaboration and acknowledgements can be found at the end of this issue.}
\address{Niels Bohr Institute, Blegdamsvej 17, 2100 Copenhagen, Denmark}

\begin{abstract} 
We report on the measurement of various flow harmonics, $v_n$, with multi-particle cumulants, and present the results from a study of the inter-correlation among different order symmetry planes $\Psi_n$ via multi-particle mixed harmonic correlations. This provides comprehensive experimental information on the fluctuating event-by-event shape of the initial conditions, which is currently among the main sources of large theoretical uncertainties in describing the evolution of the system created in heavy-ion collisions.
\end{abstract} 

\end{frontmatter} 


\section{Introduction}
\label{s:Introduction}
The properties of the produced matter in heavy-ion collisions can be experimentally studied by measuring the azimuthal anisotropy in the momentum distribution of the produced particles~\cite{Ollitrault:1992bk,Voloshin:1994mz}. Quantified by the anisotropic flow coefficients, $v_n$, and corresponding symmetry planes, $\Psi_n$, such anisotropy is expected to reflect the shape and fluctuations in the initial energy density of the collision. Fluctuations in initial geometry contribute to the even harmonic flow
and give rise to non-zero odd harmonic flow~\cite{Alver:2010gr}. Anisotropic flow harmonics are calculated as an average $v_n = \left<\cos[n(\varphi\!-\!\Psi_n)]\right>$, where $\varphi$'s are azimuthal angles of all reconstructed particles in an event. 

The contribution from fluctuations implies that $\left<v_n^k\right>$ is not the same as $\left<v_n\right>^k$, and the symmetry planes $\Psi_n$ are in general different. In these proceedings we will search for answers to the following two questions: a) What is the underlying probability density function (p.d.f.) of flow fluctuations which uniquely determine all moments $\left<v_n^k\right>$?; b) What is the relation between different symmetry planes $\Psi_n$? 

The cornerstone of our approach is the following relation for multi-particle correlations:  
\begin{equation}
\left<\cos(n_1\varphi_1+\cdots+n_k\varphi_k)\right> = v_{n_1}\cdots v_{n_k}\cos(n_1\Psi_{n_{1}}+\cdots+n_k\Psi_{n_{k}})\,,
\label{eq:mixed}
\end{equation}
derived first in~\cite{Bhalerao:2011yg}. The brackets on the LHS indicate the single-event average over all distinct tuples of $k$ particles. The last term in Eq.~(\ref{eq:mixed}) on the RHS quantifies the inter-correlations among different order symmetry planes $\Psi_n$.

One may wonder to what extent the underlying p.d.f. of $v_n$ fluctuations can be uniquely reconstructed from the measured moments $\left<v_n^k\right>$, or better from the measured multi-particle cumulants (which are less sensitive to unwanted few particle non-flow correlations). Whether or not the equivalence ``$\rm{p.d.f.} \Leftrightarrow \rm{moments} \Leftrightarrow \rm{cumulants}$" holds true was subject to debate for a long time among mathematicians. It turns out that the $2^{\rm nd}$ equivalence is always true, whilst the $1^{\rm st}$ one is not true in general. However, given the functional form of p.d.f., one can easily test whether it is uniquely determined via its moments by checking Krein-Lin conditions~\cite{Stoyanov:2006}. The most popular candidate on the theoretical frontline for the p.d.f. is a Bessel-Gaussian function, which has the remarkable property that all its multi-particle cumulants are the same~\cite{Voloshin:2007pc}. The Bessel-Gaussian function satisfies the Krein-Lin conditions, which means that it is a unique p.d.f. for which all multi-particle cumulants are the same. 

\section{Analysis details}
\label{s:Analysis}
We have used a multi-particle azimuthal correlation technique~\cite{Bilandzic:2010jr}. The analysis was performed over the dataset comprising about 30 million Pb-Pb collisions at $\sqrt{s_{\rm NN}}$ = 2.76 TeV both from the 2010 and 2011 data taking periods. Data were taken with the ALICE detector~\cite{Aamodt:2008zz}. Selected tracks of charged particles were reconstructed using the ALICE Time Projection Chamber (TPC)~\cite{Alme:2010ke}, while for centrality determination we have used the VZERO detectors. Only events with reconstructed primary vertex within 10 cm along the beam line from the nominal interaction point were selected. Charged particle tracks were required to have at least 70 reconstructed space points in the TPC and a $\left<\chi^2\right>$ per TPC cluster $\leq 4$. The maximal distance of closest approach of reconstructed tracks to primary vertex, both in longitudinal and transverse direction, was set to 3.0 cm. Event and track selection parameters were varied to estimate the systematic error of the measurement. 

\begin{figure}[ht]
\hspace{-0.25cm}
\begin{minipage}[b]{0.5\linewidth}
\centering
\includegraphics[width=1.1\textwidth]{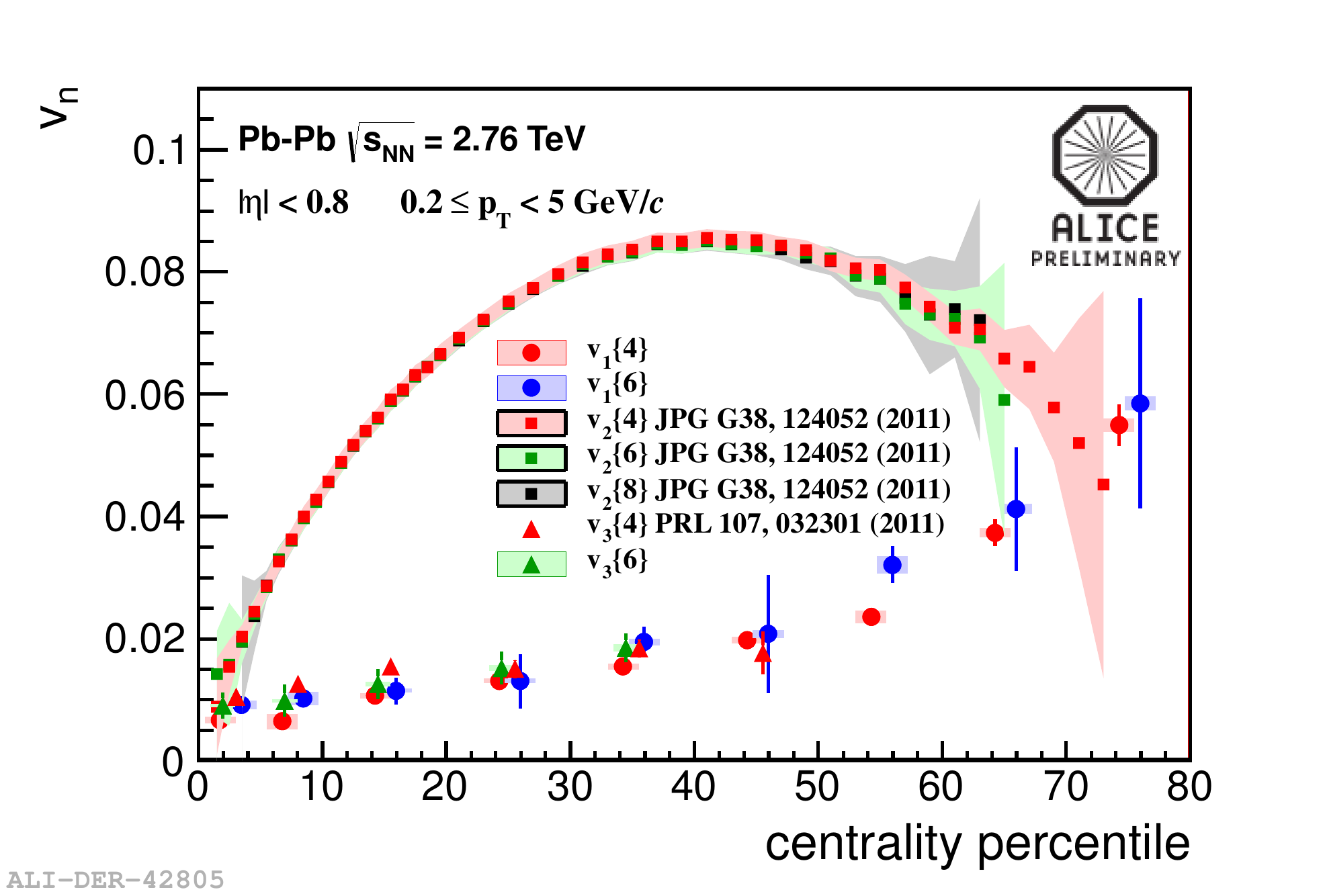}
\caption{(color online) Centrality dependence of $v_1$, $v_2$ and $v_3$ estimated with multi-particle cumulants.}
\label{fig:v146_346vsCentrality}
\end{minipage}
\hspace{0.25cm}
\begin{minipage}[b]{0.5\linewidth}
\centering
\includegraphics[width=1.1\textwidth]{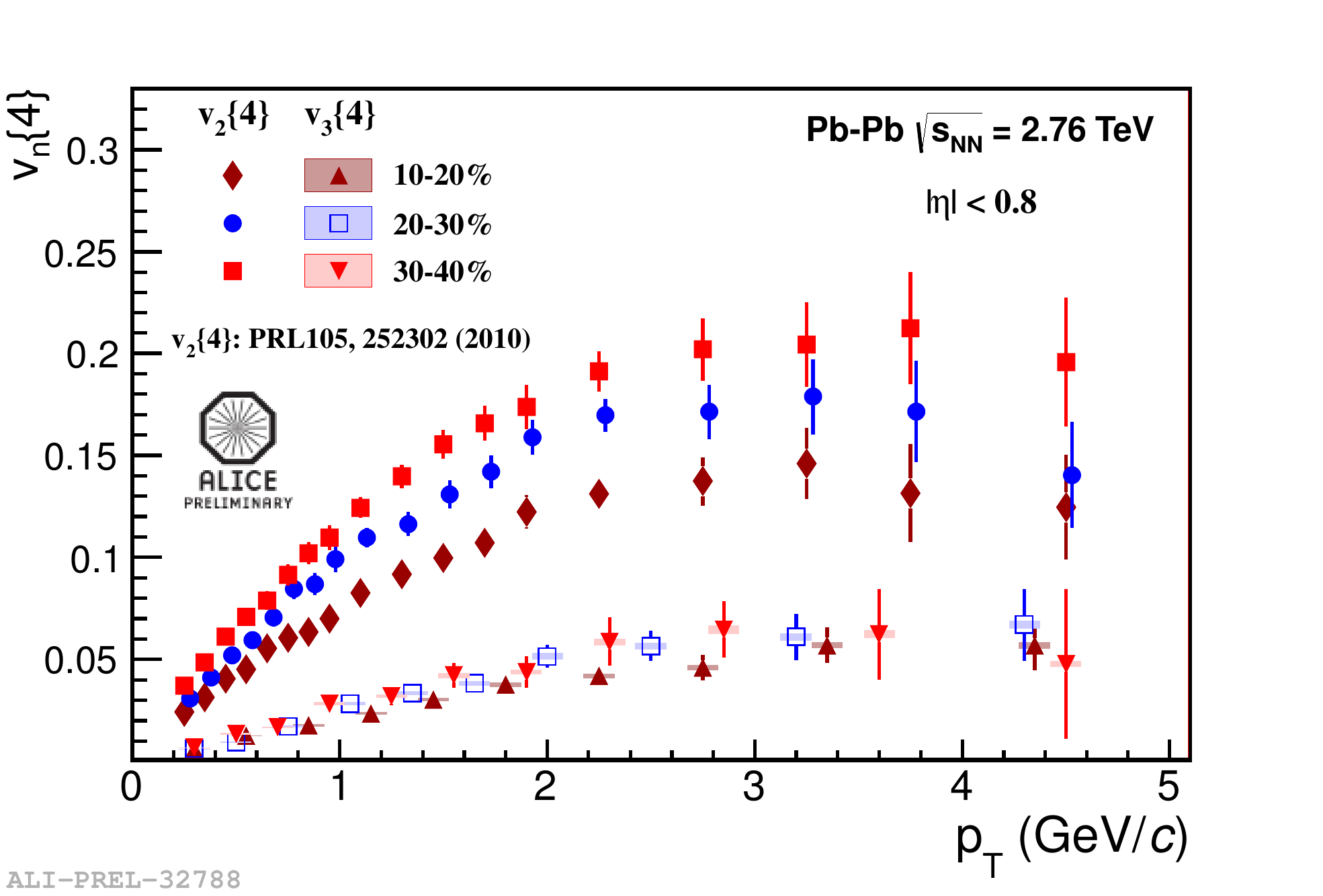}
\caption{(color online) Transverse momentum dependence of $v_2$ and $v_3$ estimated with 4-particle cumulants.}
\label{fig:v2434vsPt}
\end{minipage}
\end{figure}
%

\section{Results}
\label{s:Results}
Figure~\ref{fig:v146_346vsCentrality} shows results for $v_1$, $v_2$ and $v_3$ estimated with multi-particle cumulants (such results for $v_2$ were already available in the analysis for~\cite{Aamodt:2010pa}). We observe that for all centralities and harmonics $v_n\{4\}\simeq v_n\{6\}$. This implies that the underlying p.d.f. of flow fluctuations must have non-negligible $3^{\rm{rd}}$/higher moments when compared to the $1^{\rm{st}}$/$2^{\rm{nd}}$ moment. A Bessel-Gaussian p.d.f. is consistent with these experimental findings. 

Figure~\ref{fig:v2434vsPt} shows $v_{2}\{4\}$ and $v_{3}\{4\}$ measured as a function of transverse momentum for three different centralities, in the range $0.2 < p_{\rm T} < 5$ GeV/$c$. The coefficient $v_{3}\{4\}$ increases steadily up to $p_{\rm T} \sim$ 3 GeV/$c$ and then saturates. Similar behavior was found for $v_{2}\{4\}$,
which suggests a similar origin of both type of anisotropies. Compared to $v_2\{4\}$, $v_3\{4\}(p_{\rm T})$ changes  little with centrality. This is consistent with the widely accepted picture in which $v_{3}$ originates mainly from anisotropies stemming from event-by-event fluctuations of the initial positions of participating nucleons, while $v_2$ has an additional large contribution from almond shape geometry
of the nuclei overlap region in non-central collisions.

\begin{figure}[ht]
\hspace{-0.25cm}
\begin{minipage}[b]{0.5\linewidth}
\centering
\includegraphics[width=1.1\textwidth]{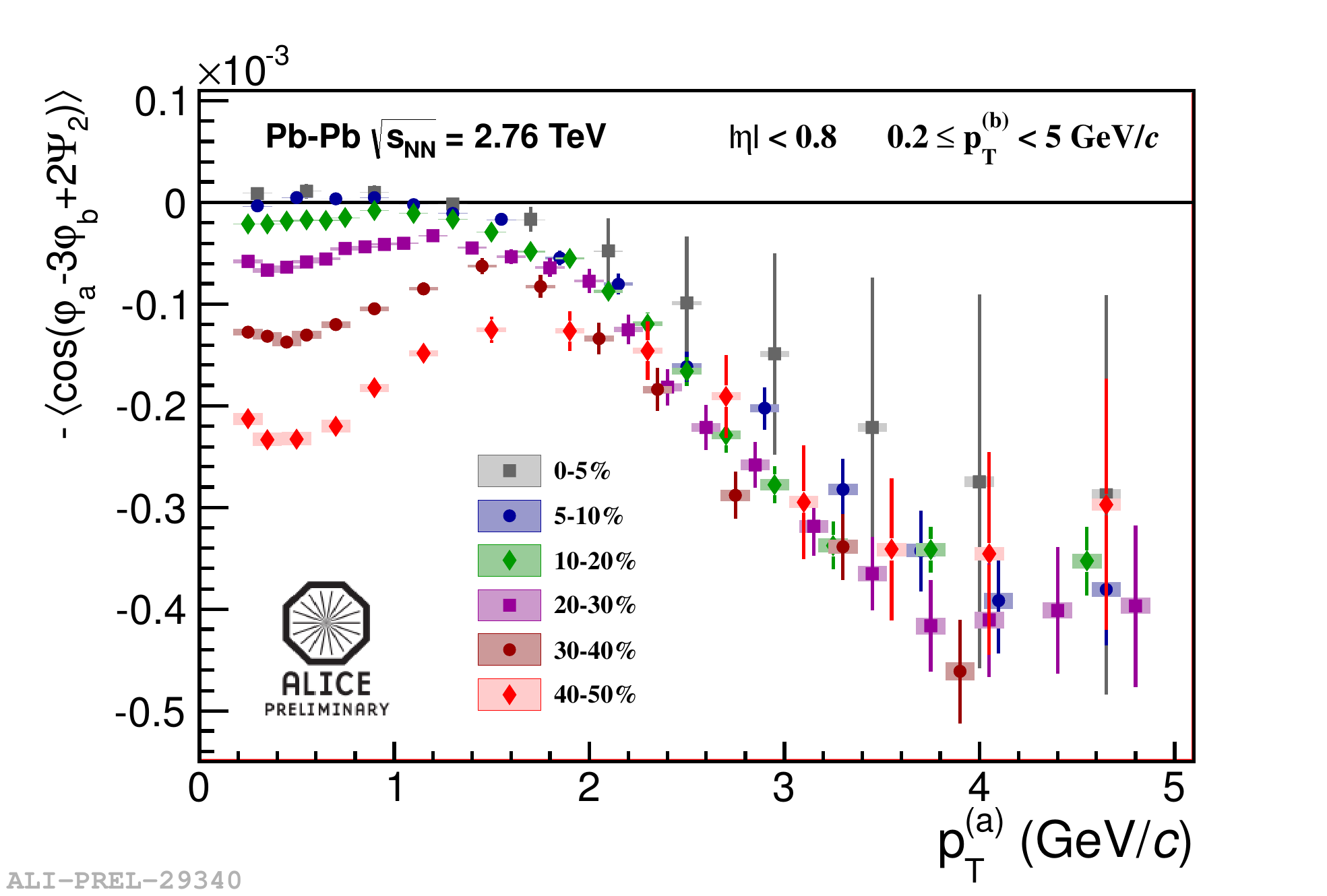}
\caption{(color online) Transverse momentum dependence of 3-particle correlator.}
\label{fig:TYvsPt_Scaled}
\end{minipage}
\hspace{0.25cm}
\begin{minipage}[b]{0.5\linewidth}
\centering
\includegraphics[width=1.1\textwidth]{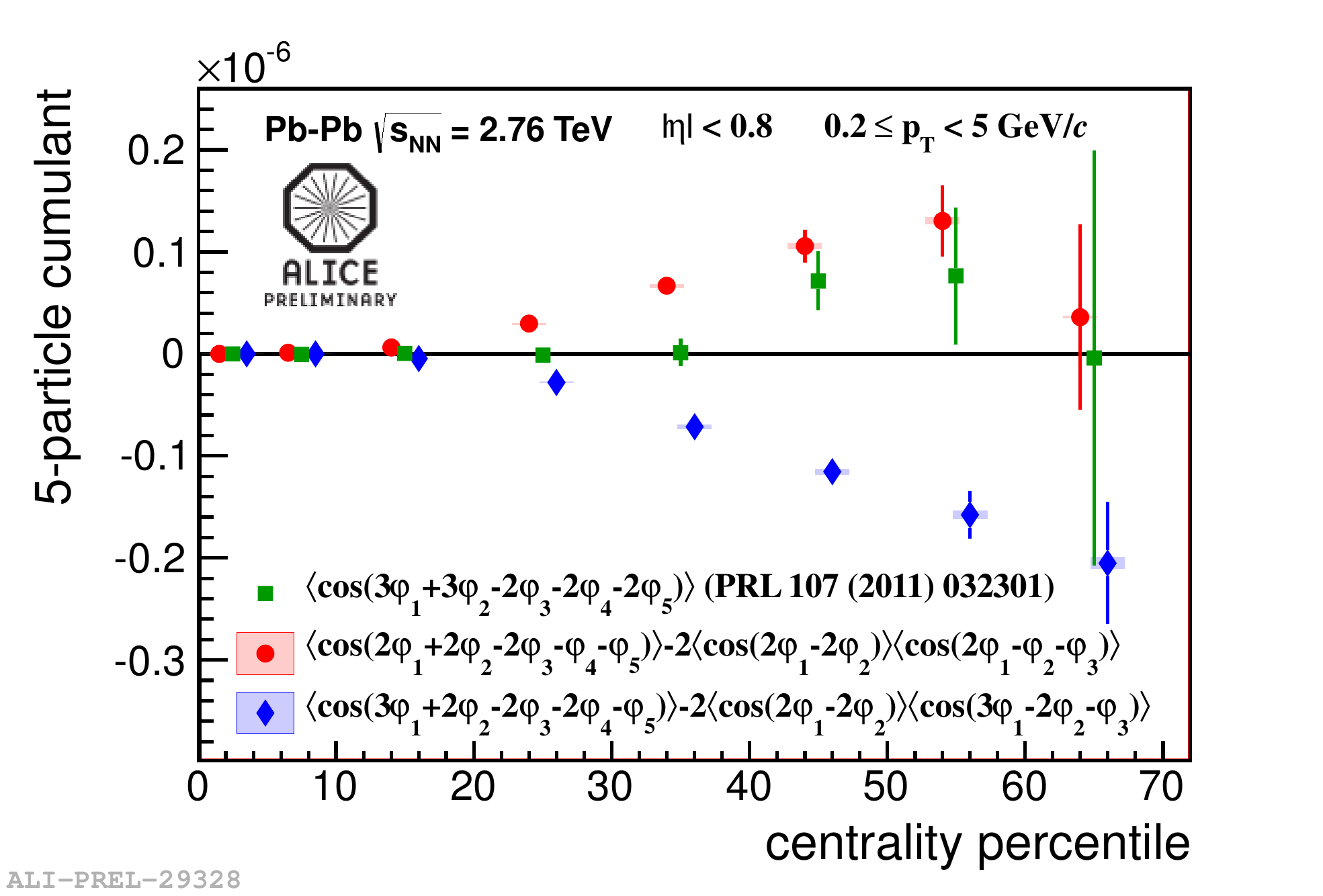}
\caption{(color online) Centrality dependence of 5-particle cumulants.}
\label{fig:QC5}
\end{minipage}
\end{figure}

Figure~\ref{fig:TYvsPt_Scaled} presents the 3-particle correlator
$\left<\cos(\varphi_a\!-\!3\varphi_b\!+\!2\Psi_2)\right>$, which was proposed first by Teaney and Yan in~\cite{Teaney:2010vd}. By making comparison with Fig.~15 from ~\cite{Teaney:2010vd}, we conclude that ideal hydrodynamic model calculations with Glauber initial conditions
can only qualitatively describe the data at low $p_{\rm T} < $ 2 GeV/$c$, while the agreement is rather poor at higher $p_{\rm T}$.

Figure~\ref{fig:QC5} presents our results obtained with 5-particle cumulants, which are sensitive to inter-correlation among different order symmetry planes $\Psi_n$, in particular to $v_3^2v_2^3\cos[6(\Psi_3\!-\!\Psi_2)]$ (green markers), $-v_2^3v_1^2\cos[2(\Psi_2\!-\!\Psi_1)]$ (red markers) and $-v_3v_2^3v_1\cos[3\Psi_3\!-\!2\Psi_2\!-\!\Psi_1]$ (blue markers). Results on 3- and 5-particle correlations show the existence of strong genuine three plane correlation $\left<\cos(\Psi_1\!-\!3\Psi_3\!+\!2\Psi_2)\right>$ among symmetry planes $\Psi_1$, $\Psi_2$ and $\Psi_3$ in mid-central collisions, although the two plane correlation $\left<\cos[6(\Psi_2-\Psi_3)]\right>$ between symmetry planes $\Psi_2$ and $\Psi_3$ is negligible in the same centralitites.  

\section{Summary}
\label{s:Summary}
We have reported in this contribution detailed results of anisotropic flow analysis which can further improve our understanding of the properties of the system produced in relativistic heavy-ion collisions. The probability density function of event-by-event flow fluctuations was studied with multi-particle cumulants, and was found to be consistent with a Bessel-Gaussian distribution. The transverse momentum dependence of triangular flow measured with 4-particle cumulants was compared to analogous previous measurements for elliptic flow, and it was demonstrated that these two observables reflect different physics, and can be used independently to constrain outputs of theoretical models. Multi-particle cumulants in mixed harmonics were reported as well, and it was shown that the theoretical framework~\cite{Teaney:2010vd} which relies on ideal hydrodynamic model calculations with the Glauber initial conditions can neither quantitatively reproduce the measurements at low $p_{\rm T} < $ 2 GeV/$c$, nor even qualitatively at higher $p_{\rm T}$.

\end{document}